\documentstyle[epsf,graphics,epsfig]{mn}

\newcommand{\etal}{{\it et~al.}}
\newcommand{\msun}{\thinspace\hbox{$M_{\odot}$}}

\title[The build-up of halos within Press-Schechter theory]
{The build-up of halos within Press-Schechter theory}

\author[Will J. Percival]{Will J. Percival\\
  Institute for Astronomy, University of Edinburgh, 
  Royal Observatory, Blackford Hill, Edinburgh EH9 3HJ, U.K.\\}

\date{Submitted for publication in MNRAS}
\begin{document}
\maketitle

\begin{abstract}
Modelling the build-up of halos is important for linking the formation
of galaxies with cosmological models. A simple model of halo growth is
provided by Press-Schechter (PS) theory, where the initial field of
density fluctuations is smoothed using spherically symmetric filters
centred on a given position to obtain information about the likelihood
of later collapse on varying scales. In this paper the predicted halo
mass growth is compared for three filter shapes: Gaussian, top-hat and
sharp $k$-space. Preliminary work is also presented analysing the
build-up of halos within numerical simulations using a
friends-of-friends group finder. The best-fit to the simulation mass
function was obtained using PS theory with a top-hat filter. By
comparing both the backwards conditional mass function, which gives
the distribution of halo progenitors, and the distribution of halo
mergers in time, the build-up of halos in the simulations is shown to
be better fitted by PS theory with a sharp $k$-space filter. This
strengthens previous work, which also found the build-up of halos in
simulations to be well matched to PS theory with a sharp $k$-space
filter by providing a direct comparison of different filters and by
extending the statistical tools used to analyse halo mass growth. The
usefulness of this work is illustrated by showing that the
cosmological evolution in the proportion of halos that have undergone
recent merger is predicted to be independent of mass and power
spectrum and to only depend upon cosmology. Recent results from
observations of field galaxies are shown to match the evolution
expected, but are not sufficiently accurate to usefully distinguish
between cosmological parameters.
\end{abstract}


\section{Introduction}

Deep surveys are beginning to probe the morphological evolution of
non-active field galaxies out to redshifts where cosmological changes
in the population as a whole become apparent. The evolution of such
galaxies is expected to be driven by the hierarchical build-up of the
dark matter halos in which the galaxies reside, with mergers between
halos of approximately equal size being particularly important. Such
mergers can lead to the disruption of galaxy structure producing
observable signatures. Consequently, it is important to be able to
model the growth of halos in sufficient detail to determine the
relative frequency of mergers.

Press-Schechter (PS) theory \cite{ps} provides a simple analytic model
for the build-up of halos. In this theory, the initial field of
density fluctuations is smoothed on varying scales around a given
position to obtain information about the probability of later
collapse. The distribution of overdensities as a function of filter
size is called a trajectory. If filter size and halo mass are linked,
each trajectory contains information about the mass of halo within
which a test-particle associated with that trajectory and position in
the linear density field resides. Linking perturbation amplitude with
the epoch at which fluctuations undergo non-linear collapse, and
assuming that collapse occurs on the largest scale for which there is
sufficient excess overdensity provides a complete model for halo mass
growth.

The standard use of PS theory is to calculate the mass function
$f(M|t)\,dM$, the distribution of mass in isolated halos at a given
epoch \cite{ps,peacock,bond}. For the sharp $k$-space filter, it is
possible to obtain an analytic formula, although this is not true for
more general filters. A numerical method for generating trajectories
for general spherically symmetric filters was given by Bond \etal\
\shortcite{bond} and is summarized in Section~\ref{sec:ps}. The effect
of the choice of filter on the resulting mass function is considered
in Section~\ref{sec:mf}, where the top-hat filter is shown to provide
the best match to recent N-body simulation results. The distribution
of times at which halos of a given mass are created can also be
determined by PS theory \cite{ev1}, and is discussed for different
filters in Section~\ref{sec:cr}.

It is possible to use PS theory to follow the build-up of halos that
have mass $M$ at time $t$ by considering the behaviour of trajectories
that pass through this location. This is often called `extended PS
theory' and leads to conditional analogues of the mass function and
halo creation rate \cite{bower,bond,lc93,ev1}. Extended PS theory is
often used to estimate halo merger trees, which are in turn used in
semi-analytic models of galaxy formation
\cite{kauffmann_ht,sheth_ht,somerville_ht}. The distribution of
progenitor masses at an earlier epoch is termed the backwards
conditional mass function, while the distribution of halo masses in
which a particle might find itself at a later time is termed the
forward conditional mass function. Lacey \& Cole \shortcite{lc94}
showed that the forward conditional mass function from simulations
with power law spectra appears to be well fitted by a variant of PS
theory. In this paper this work is extended to determine the effect of
filter choice on the backwards conditional mass function and the
predicted frequency of merger events as a mechanism for the build-up
of dark matter halos. The backwards mass function is analysed in
Section~\ref{sec:cmf}. Starting with the premise that mergers between
halos correspond to jumps in the mass associated with a single
trajectory in PS theory, the relative frequency of mergers are
compared for different filters in Section~\ref{sec:merg}.

The usefulness of this analysis is demonstrated in
Section~\ref{sec:obs}, where the proportion of halos that have
undergone recent merger is calculated. This is linked to the field
galaxy merger rate, and recent observational results are shown to be
consistent with a variety of cosmological models. Further observations
could, in principle, lead to estimates of $\Omega_m$ and $\Omega_v$.

\section{The principles of Press-Schechter theory}  \label{sec:ps}

\begin{figure*}
  \setlength{\epsfxsize}{\textwidth} \centerline{\epsfbox{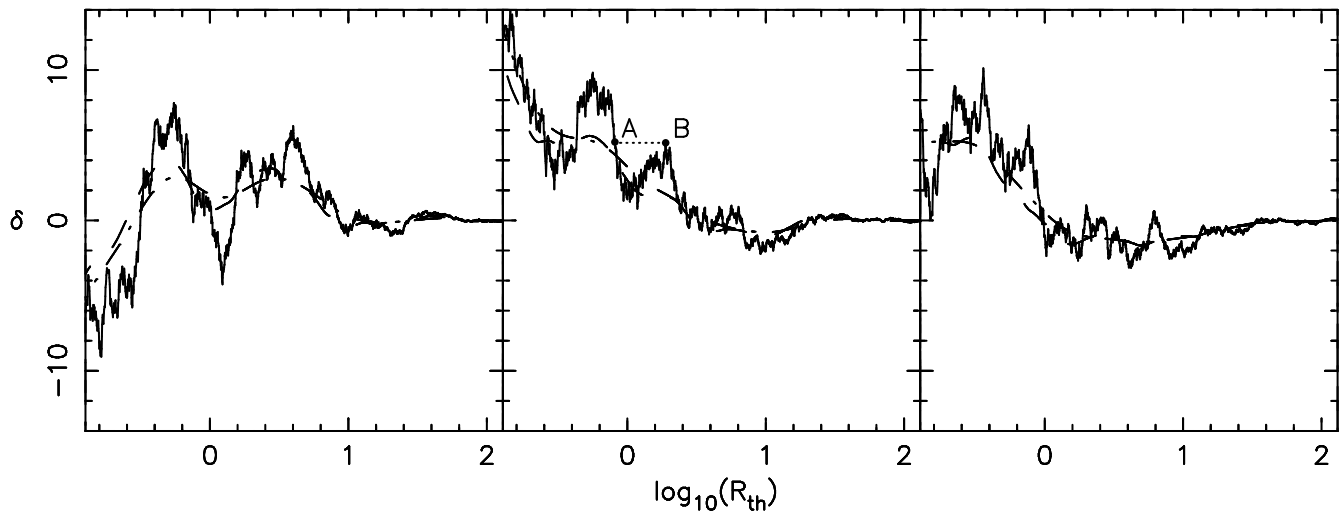}}
  \caption{Trajectories are plotted calculated by Langevin integration
  for three independent positions in a Gaussian random field with CDM
  power spectrum. For each position, an estimate of the density field
  was made integrated over $2^{16}$ contiguous shells equally spaced
  in $\ln k$. Each Monte-Carlo realisation of the field was then
  filtered with sharp $k$-space (solid line), top-hat (dashed line)
  and Gaussian (dot-dash line) filters at $2^{12}$ radii, equally
  spaced in $\log\sigma_M^2$. The top-hat and Gaussian filters provide
  smoother trajectories than the sharp $k$-space filter for the same
  field position although, as expected, the overall behaviour is
  similar for all filters. Trajectories are matched as described in
  Section~\ref{sec:filters}. The mass associated with the sharp
  $k$-space trajectory shown in the centre panel will undergo a jump
  from a value corresponding to $R_A$ to a value corresponding to
  $R_B$ at a time when $\delta_c=\delta_{AB}$, the value along the
  dotted line.} \label{fig:traj}
\end{figure*}

Press-Schechter (PS) theory is based on two simple ideas. First, the
field of initial density fluctuations is filtered to determine the
amplitude of the perturbations on different scales. Second, the
perturbation amplitudes are linked to their collapse time or lack of
it, thereby providing a complete statistical model for the build-up of
structure in the Universe. Three filters are compared in this paper,
defined in $k$-space by
\begin{eqnarray}
  \lefteqn{\mbox{Sharp $k$-space}} \nonumber \\
    &\tilde{W_{ks}}(k;R_{ks})=&\vartheta\left(1-kR_{ks}\right)
       \label{eq:fk} \\
    & & \nonumber  \\
  \lefteqn{\mbox{Top-Hat}} \nonumber \\
    &\tilde{W_{th}}(k;R_{th})=&3\frac{\sin(kR_{th})-(kR_{th})\cos(kR_{th})}
      {(kR_{th})^3} \label{eq:fth} \\
    & & \nonumber  \\
  \lefteqn{\mbox{Gaussian}} \nonumber \\
    &\tilde{W_g}(k;R_g)=&\exp\left(\frac{-R_g^2k^2}{2}\right) 
      \label{eq:fg}
\end{eqnarray}
where $\vartheta(x)$ is the Heaviside step function, $\vartheta(x)=1$
for $x\geq0$ and $\vartheta(x)=0$ otherwise. The mass variance,
$\sigma_M^2$ inside a volume $V$ whose shape and size are dependent on
the filter is given by
\begin{equation}
  \sigma_M^2=\frac{1}{2\pi^2}\int_0^{\infty}P(k)
    \tilde{W}^2(k;R)k^2\,dk,
  \label{eq:sigmasq}
\end{equation}
where P(k) is the power spectrum of the initial density perturbations,
and is often simply referred to as the power spectrum. Unless stated
otherwise, the standard CDM power spectrum of Bond \etal\
\shortcite{bond} has been assumed with
$\Gamma\equiv\Omega_mh=0.5$. The set of overdensities corresponding to
different $R$ for a given position in the linear density field (this
can also be thought of as corresponding to a test-particle or small
mass element) is called a trajectory.

Examples of trajectories created for the filters defined by
Equations~\ref{eq:fk},~\ref{eq:fth}~\&~\ref{eq:fg} are presented in
Fig.~\ref{fig:traj}. These trajectories were calculated using Langevin
integration as described by Bond \etal\ \shortcite{bond}. First, a
Monte-Carlo realisation of a Gaussian random field was created,
integrated over contiguous spherical shells in $k$-space. The
amplitudes of the field integrated over non-overlapping shells are
independent and, for thin shells from $\ln k\to \ln k+\delta\ln k$,
are given by a Gaussian random variable with variance
\begin{equation}
  \sigma_{\rm shell}^2=\frac{1}{2\pi^2}k^3P(k)\delta\ln k.
\end{equation}
For the trajectories presented in
Sections~\ref{sec:mf}~\&~\ref{sec:cr}, the field was determined around
a given position for $2^{12}$ shells equally spaced in $-7<\ln
k<6$. Trajectories were then created by summing over these shells,
weighted by the filter window function for $2^8$ radii equally spaced
in $\log\sigma_M^2$. As a check that the density field is adequately
sampled, results are presented in
Sections~\ref{sec:mf}~\&~\ref{sec:cr} for the sharp $k$-space filter,
for which the result is known analytically. Tests have also shown that
increasing or decreasing the sampling does not significantly alter the
results. A detailed analysis of the sampling required to calculate the
backwards conditional mass function is given in Section~\ref{sec:cmf}.

In order to provide a complete description of the hierarchical
build-up of mass, the epoch of collapse must be linked with the
overdensity. In fact, the link between the critical overdensity
$\delta_c$ and the collapse epoch is predicted to be independent of
mass for the simple spherical top-hat collapse model (Percival, Miller
\& Peacock 2000), the model adopted in standard PS theory. The mass of
halo surrounding a given position at time $t$ is then given by the
first, or largest mass, upcrossing of a horizontal barrier
$\delta=\delta_c(t)$ by the trajectory.

\section{Parameters in PS theory}  \label{sec:filters}

This Section discusses the free parameters in Press-Schechter theory,
and outlines the conventions used in this paper.

The choice of filter was presented as a single free parameter in
Section~\ref{sec:ps}. However, because PS theory only predicts an
analytic form for the unconditional and conditional mass functions for
the sharp $k$-space filter, many authors have in the past used these
analytic functions and have calculated $\sigma_M^2$ using a different
filter. The choice of filter used to calculate $\sigma_M^2$ becomes an
additional free parameter in PS theory (with the choice of filter to
determine the mass function being a different free parameter). For
instance, both Kauffmann \& White (1993) and Somerville \& Kolatt
(1999) use the spatial top-hat filter to calculate $\sigma_M^2$, and
the sharp $k$-space filter to determine the mass functions. Similarly,
Lacey \& Cole (1994), allowed the choice of filter used to calculate
$\sigma_M^2$ to be different to the sharp $k$-space filter used to
determine the mass function. The effect of this switch is equivalent
to using the sharp $k$-space filter throughout with a slightly
modified power spectrum. In this paper, the same filter is used to
determine both the mass function and $\sigma_M^2$.

Another parameter that is often allowed to vary is the critical
overdensity for collapse, $\delta_c$, with authors tuning this
parameter to fit simulations (Lacey \& Cole 1994), or using the
ellipsoidal collapse model (Sheth, Mo \& Tormen 2001). For simplicity,
only the spherical top-hat collapse model is considered in this paper.

\begin{figure}
  \setlength{\epsfxsize}{0.75\columnwidth}
  \centerline{\epsfbox{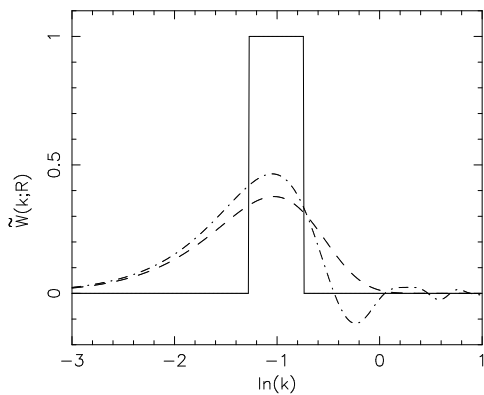}} 

  \caption{The transitional weighting applied to the $k$-shells for
  one step in a trajectory for the sharp $k$-space (solid line),
  top-hat (dashed line) and Gaussian (dot-dash line) filters for a
  change in mass of between $10^{14}$ and $5\times10^{14}$\,\msun. The
  filters have been matched so that the transitional weighting peaks
  at the same value of $\ln k$, as described in
  Section~\ref{sec:filters}.}  \label{fig:filters}
\end{figure}

\begin{figure*}
  \setlength{\epsfxsize}{\textwidth} \epsfbox{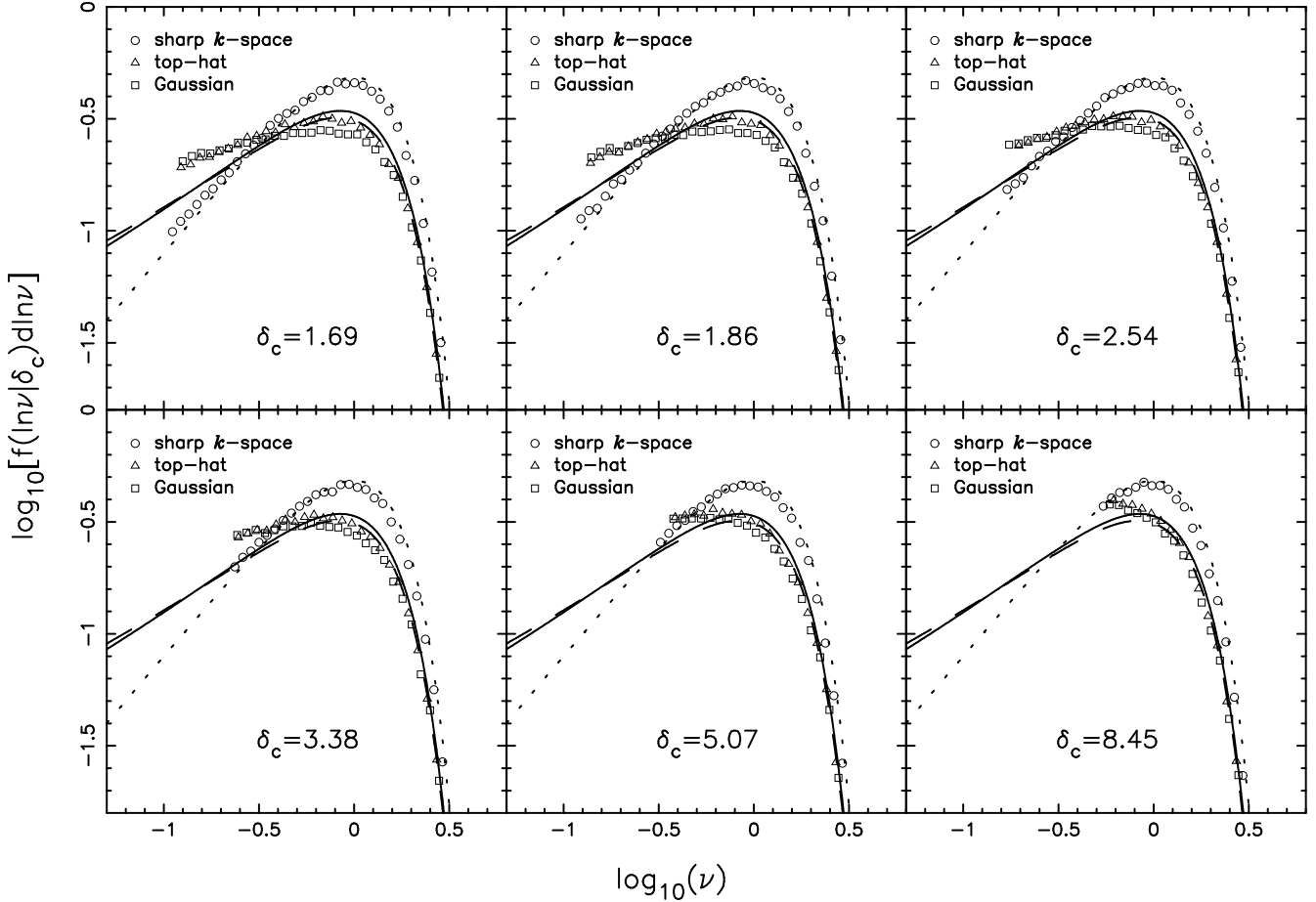} 

  \caption{The halo mass function is presented for each filter as a
  function of $\nu\equiv\delta_c/\sigma_M$ for six values of
  $\delta_c$, calculated from $10^6$ Monte-Carlo trajectories.  The
  six values of $\delta_c$ correspond to redshifts of 0.0, 0.1, 0.5,
  1.0, 2.0 and 4.0 for an Einstein-de Sitter cosmology. The prediction
  of PS theory for a sharp $k$-space filter is also plotted (dotted
  line), and is a close match to the corresponding numerical data. The
  slight deviation is caused by differing sampling of the field and
  trajectory. Because of the filter shapes, such a difference will
  affect results for the top-hat \& Gaussian filters less. Results for
  different filters were matched as described in
  Section~\ref{sec:filters}. The dashed line shows the fit to N-body
  simulation results of Sheth \& Tormen \protect\shortcite{sheth}, and
  the solid line the fit of Percival \etal\ \protect\shortcite{ev2}.}
  \label{fig:mf}
\end{figure*}

In order to compare results calculated using different filters, the
radii of the different filters need to be matched. In each step of a
trajectory, the power spectrum is convolved with a transitional
weighting function. Fig.~\ref{fig:filters} shows an example of the
transitional weighting applied by the three filters considered in this
paper. Matching the radii of the different filters effectively matches
the weightings applied to the power spectrum. A common method is to
integrate each filter over all real space to calculate its `volume of
influence' and the average mass enclosed. Equating volumes leads to
\begin{equation}
  R_{ks}=\left(\frac{9\pi}{2}\right)^{-1/3}R_{th}\simeq0.414R_{th}, 
\end{equation}
\begin{equation}
  R_g=\left(\frac{4\pi}{3}\right)^{1/3}\frac{1}{(2\pi)^{1/2}}R_{th}
    \simeq0.643R_{th}.
\end{equation}
For the differentiable filters it is also possible to match filter
radii to second order in the expansion around $k=0$. This requires
$R_g=1/\sqrt{5}R_{th}\simeq0.447R_{th}$.

Each point in a trajectory is calculated by integrating over
$k$-shells, and matching different filters simply alters the region of
the power spectrum probed. It is possible to directly match the peak
value of $k$ probed at a given step in a trajectory. For the
non-differentiable sharp $k$-space filter, $k_{\rm max}=1/R_{ks}$. For
the Gaussian filter the first solution to the differential equation
$d^2\tilde{W}/dRdk=0$ is given by $k_{\rm max}=\sqrt{2}/R_{g}$ and for
the top-hat filter solving this equation numerically gives $k_{\rm
max}\simeq3.342/R_{th}$. Combining these equations gives
\begin{equation}
  R_{ks}\simeq0.299R_{th}, \hspace{1cm} R_g\simeq0.423R_{th}.
  \label{eq:Rrel}
\end{equation}
The differential weighting applied in a particular trajectory `step'
by each filter for this choice of matching filters is shown in
Fig.~\ref{fig:filters}.

In order to highlight the effect of filter choice on halo growth, it
seems prudent to ensure that the same region of the power spectrum is
probed at each trajectory step. Because of this, where appropriate,
the filter radii have been matched using Equation~\ref{eq:Rrel}.
Having adopted this convention, the halo mass $M$ was defined as
\begin{equation}
M=\frac{4\pi}{3}\bar{\rho}R_{th}^3,
\end{equation}
where $\bar{\rho}$ is the mean density of the universe. Because the
mass variance, $\sigma_M^2$ is dependent on filter choice
(Equation~\ref{eq:sigmasq}), comparing results as a function of
$\sigma_M^2$ implicitly matches filter radii. In this case, the
relationship between the radii of different filters is dependent on
the power spectrum. Filter radii are only matched using
Equation~\ref{eq:Rrel} when the halo mass is required.

\section{The halo mass function}  \label{sec:mf}

\begin{figure*}
  \setlength{\epsfxsize}{\textwidth} \epsfbox{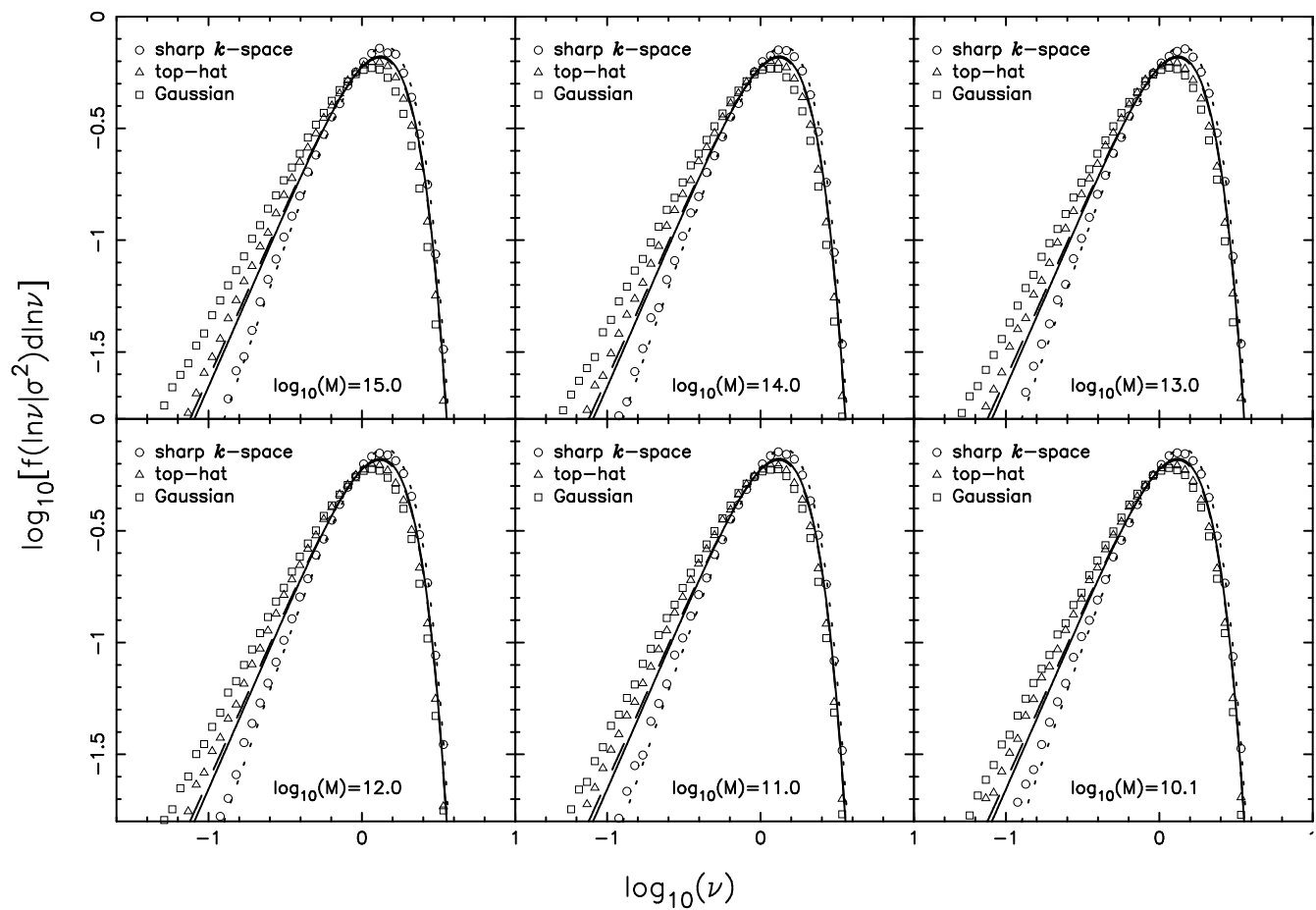}

  \caption{The distribution of halo creation times are plotted for six
  halo masses as a function of $\nu\equiv\delta_c/\sigma_M$. These
  data were calculated from $10^6$ Monte-Carlo trajectories created as
  described in Section~\ref{sec:ps}. The lines are as in
  Fig.~\ref{fig:mf}, converted into creation rates by multiplying by
  $\nu$ \protect\cite{ev2}. The Monte-Carlo data have been normalized
  to the model of Sheth \& Tormen \protect\shortcite{sheth}.}
  \label{fig:cr}

\end{figure*}

The mass function $f(M|t)\,dM$, gives the distribution of mass in
isolated halos at a given epoch, and is related to the number density
of halos by $\bar{\rho} f(M|t)\,dM=Mn(M)\,dM$. The mass function can
be calculated analytically for a sharp $k$-space filter because the
trajectories are Brownian random walks and therefore have special
symmetries \cite{ps,peacock,bond}. Defining
\begin{equation}
  \nu\equiv\frac{\delta_c}{\sigma_M},  \label{eq:nu}
\end{equation}
the PS mass function can be written as
\begin{equation}
  f(\ln\nu|t)\,d\ln\nu = \sqrt{\frac{2}{\pi}}\nu e^{-\nu^2/2}\,d\ln\nu.
  \label{eq:psmf}
\end{equation}
Note that when the mass function is written in this way, its form is
independent of epoch. It is not possible to obtain corresponding
analytic formulae for top-hat or Gaussian filtering \cite{bond}.

By creating an ensemble of $10^6$ trajectories for each of the filters
given by Equations~\ref{eq:fk},~\ref{eq:fth}~\&~\ref{eq:fg} using the
method described in Section~\ref{sec:ps}, the mass function has been
estimated at six epochs. These data are presented in Fig.~\ref{fig:mf}
as a function of $\nu$.

Recent work analysing the mass function predicted by numerical
simulations has suggested that, although the mass function has a
universal form when written as a function of $\nu$, this form is
altered from that of Equation~\ref{eq:psmf}
\cite{sheth,ev2,jenkins2000}. Instead, Sheth \& Tormen
\shortcite{sheth} found that a good fit to the N-body data was given
by
\begin{equation}
  f(\ln\nu|t)\,d\ln\nu 
    = A\sqrt{\frac{2}{\pi}}\left(1+\frac{1}{\nu'^{2p}}\right)\nu'e^{-\nu'^2/2}
    \,d\ln\nu,
  \label{eq:sheth}
\end{equation}
where $\nu'=a^{1/2}\nu$ and $a$~\&~$p$ are parameters. $A$ is
determined by requiring that the probability density function is
correctly normalized. Sheth \& Tormen found best fit parameters
$a=0.707$ and $p=0.3$ for their simulations and group finding
algorithm. The standard PS multiplicity function has $a=1$, $p=0$ and
$A=1/2$. In Percival \etal\ \shortcite{ev2}, a small tweak was
required for the parameters in order to fit the data giving $a=0.774$
\&~$p=0.274$. These fits to the mass function are presented in
Fig.~\ref{fig:mf}.

The mass function calculated from trajectories determined for the
sharp $k$-space has close to its predicted analytic behaviour given by
Equation~\ref{eq:psmf}, showing that the Monte-Carlo procedure works
for this filter. The smoother trajectories calculated for the top-hat
and Gaussian filters produce mass functions more like that observed in
numerical simulations, with fewer high mass halos than predicted by
Equation~\ref{eq:psmf}. This match is not perfect, and the smoother
filters predict more low mass halos than Equation~\ref{eq:sheth}.
 
\section{The halo creation rate}  \label{sec:cr}

\begin{figure*}
  \setlength{\epsfxsize}{\textwidth} \epsfbox{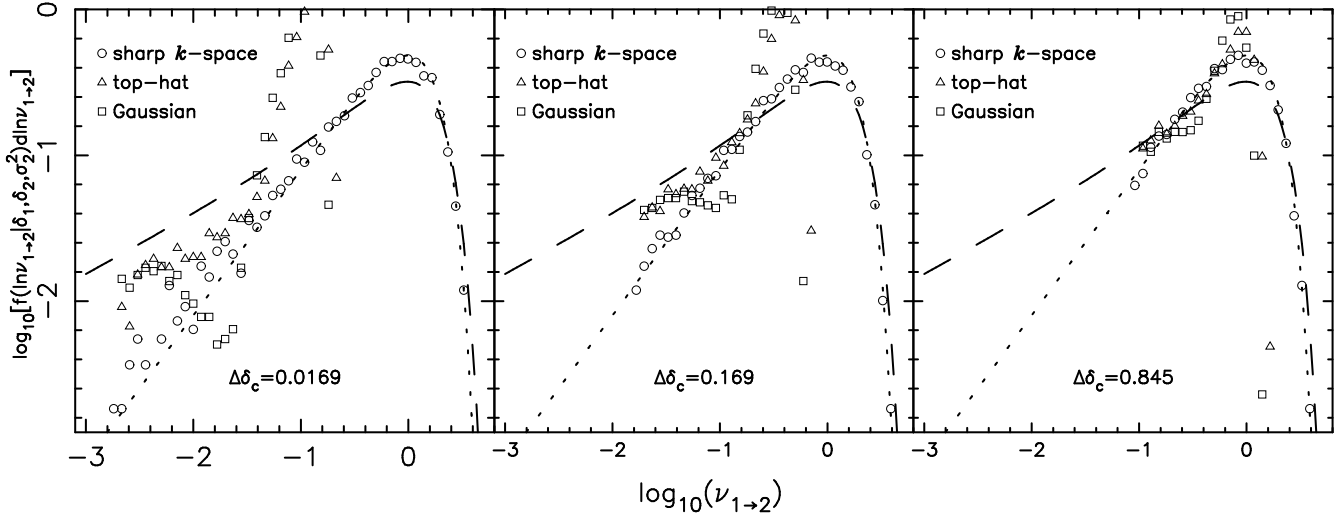} \caption{The
  backwards conditional mass function calculated for PS theory with
  three filters is presented as a function of $\nu_{1\to2}$, for
  $\sigma_2^2$ equivalent to a mass of $10^{14}$\msun. All
  trajectories that predict a halo of mass $10^{14}$\msun\ at
  $\delta_2>1.69$ were included in this analysis. The dotted line
  shows the analytic prediction of PS theory with a sharp $k$-space
  filter, and the dashed line shows the fit to the mass function of
  Sheth \& Tormen \protect\shortcite{sheth}, presented as if its form
  is invariant to the change $\nu\to\nu_{1\to2}$.}
  \label{fig:psf_cmf}
\end{figure*}

The halo creation rate $f(t|M)\,dt$ for halos of mass $M$ is given by
the distribution of first upcrossing times of a barrier at the filter
radius corresponding to $M$. The distribution of halo creation times
determined for Gaussian, top-hat and sharp $k$-space filters are
plotted in Fig.~\ref{fig:cr} for the subset of Monte-Carlo
trajectories that predict the creation of a halo with one of six
masses. Each trajectory does not necessarily have a first upcrossing
at a particular mass because the mass can undergo finite `jumps' (see
Section~\ref{sec:merg}). Consequently, the total number of creation
events recorded from a fixed number of trajectories depends on the
filter chosen. As the sharp $k$-space filter provides a `rough'
trajectory compared to top-hat and Gaussian, there are more mass jumps
and less creation events for this filter at a given final halo
mass. In fact, it is only because of the discrete nature of these
trajectories that any creation events at a particular mass are found.

It follows from the Bayesian argument presented in Percival \etal\
\shortcite{ev2} that
\begin{equation}
  \nu f(\ln\nu|t)\,d\ln\nu \propto f(\ln\nu|M)\,d\ln\nu.
\end{equation}
In the first of these probability density functions $\nu$ is
considered to be a function of mass, while in the second it is
considered to be a function of time. The model mass functions plotted
in Fig.~\ref{fig:mf} have been converted to creation rates by
multiplying by $\nu$ and renormalising, and are compared with the
creation rate calculated from the Monte-Carlo trajectories in
Fig.~\ref{fig:cr}.

The results are similar to those of Fig.~\ref{fig:mf}, with the
Monte-Carlo creation rate calculated for the sharp $k$-space filter
closely following the expected form. The creation rates calculated
using the fit to the N-body simulations (Equation~\ref{eq:sheth}) are
closer to PS theory with the top-hat and Gaussian filters although, as
for the mass function, the match is not perfect. Comparing
Figs.~\ref{fig:mf}~\&~\ref{fig:cr} shows that the Bayesian link
between the halo creation rate and mass function is closely followed
by the Monte-Carlo data.

\section{The backwards conditional mass function}  \label{sec:cmf}

\subsection{PS theory}

The subset of trajectories that pass through a selected position
determine the probability density functions for the past and future
masses of that halo. As for the mass function, analytic formulae are
possible for the sharp $k$-space filter \cite{bower,bond,lc93},
whereas for top-hat and Gaussian filters numerical techniques must be
used. The conditional analogue of $\nu$ is defined as
\begin{equation}
  \nu_{1\to2}\equiv\frac{\delta_1-\delta_2}{\sqrt{\sigma_1^2-\sigma_2^2}},
\end{equation}
where $\sigma_2^2<\sigma_1^2$ ($M_2>M_1$) and $\delta_2<\delta_1$
($t_2>t_1$), and the subscripts $M$ and $c$ have been omitted for
simplicity. For the backwards conditional mass function, $\sigma_2^2$
and $\delta_2$ (and $\delta_1$) are known and $\nu_{1\to2}$ is
considered to be a function of $\sigma_1^2$. For the sharp $k$-space
filter
\begin{displaymath}
  f(\ln\nu_{1\to2}|\delta_1,\delta_2,\sigma_2^2)\,d\ln\nu_{1\to2}
\end{displaymath}
\begin{equation}
  \hspace{1cm}
    =\sqrt{\frac{2}{\pi}}\nu_{1\to2}
    e^{-\nu_{1\to2}^2/2}\,d\ln\nu_{1\to2},
  \label{eq:cmf}
\end{equation}
which has the same form as Equation~\ref{eq:psmf}. This is expected
because the form of the trajectories is invariant to a linear shift in
$\delta$ or $\sigma^2$.

For the top-hat and Gaussian filters, the equivalent distribution can
be calculated by analysing numerical realisations of trajectories. In
order that all upcrossings at small
$\Delta\delta_c\equiv(\delta_2-\delta_1)$ were realized, the
trajectories used for this analysis were sampled at $2^{10}$ steps in
$R$, equally spaced in $-6<\log_{10}(\sigma_2^2-\sigma_1^2)<2$. For
the sharp $k$-space filter, only a single $k$-shell covering each
interval in $R$ is required. The other two filters have a broader
shape and each step in a trajectory will depend on a larger region of
$k$-space. Two $k$-shells within each interval in $R$ and $2^7$ shells
either side of the minimum and maximum radii, equally spaced in $\ln
k$ were found to sufficiently sample $k$-space.

The backwards conditional mass functions resulting from these
trajectories are presented in Fig.~\ref{fig:psf_cmf} for three values
of $\Delta\delta_c$. In order to match the N-body results (for an
Einstein-de Sitter cosmology), only trajectories that predict a halo
of mass $10^{14}$\msun\ at $\delta_1>1.69$ have been counted. The
smoother trajectories calculated using top-hat and Gaussian filters
predict low mass progenitors, particularly for small $\Delta\delta_c$.

\begin{figure*}
  \setlength{\epsfxsize}{\textwidth} \epsfbox{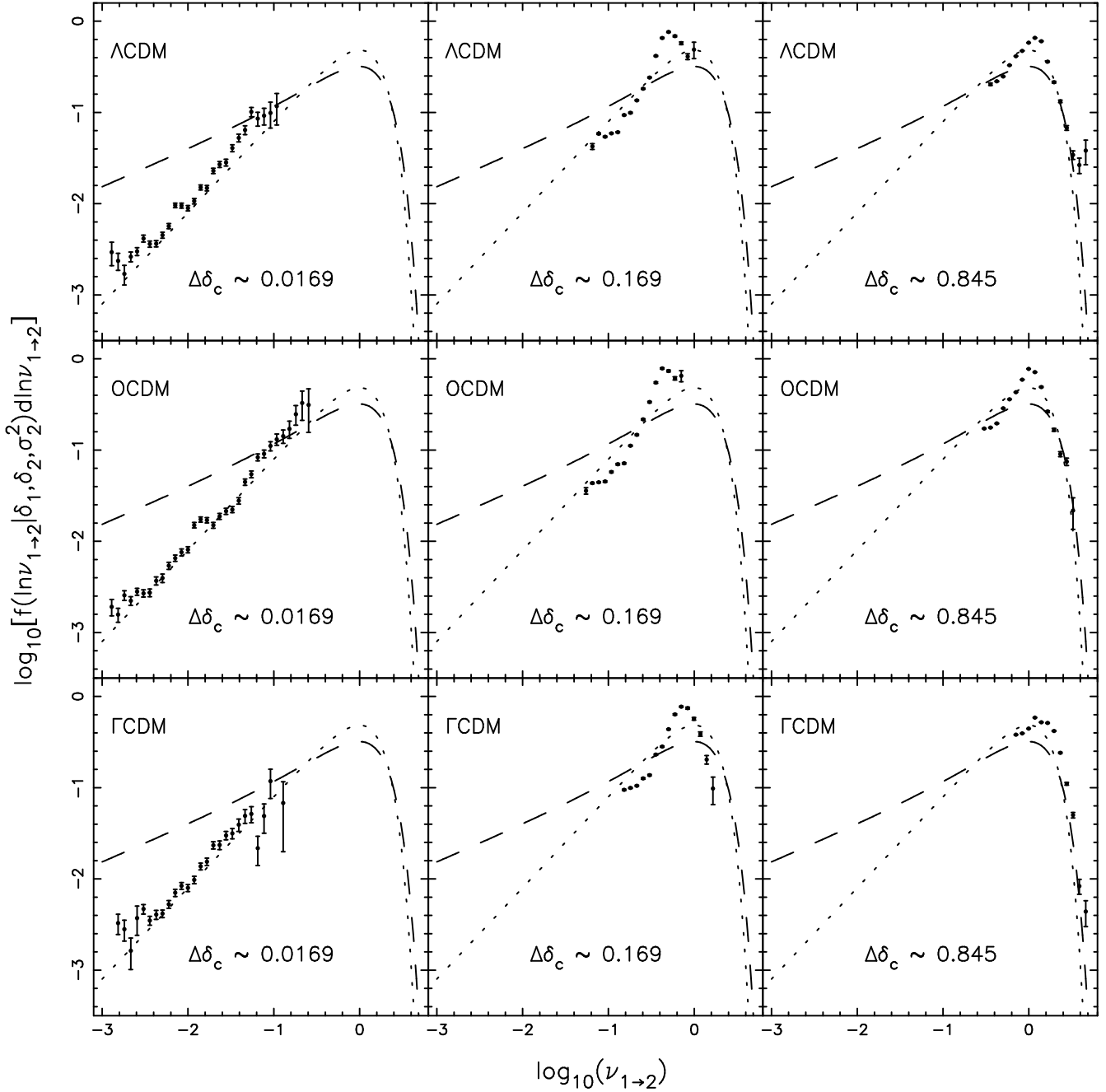} \caption{The
  backward conditional mass function calculated from three N-body
  simulations (See text for details). The left column shows the
  backwards conditional mass function from consecutive output from the
  simulation, while the centre column shows the backward conditional
  mass function for $\Delta\delta_c\sim0.169$, and the right column
  $\Delta\delta_c\sim0.845$. The rows show results from different
  simulations. Lines are plotted as in Fig.~\ref{fig:psf_cmf}.}
  \label{fig:nb_cmf}
\end{figure*}

\subsection{Numerical Simulations}

Three numerical simulations, completed using the Hydra N-body,
hydrodynamics code \cite{couchman}, have been analysed in order to
estimate the backwards conditional mass function for halos of mass
$\sim10^{14}$\msun. Details of the cosmologies adopted for these
simulations are summarized in Table~\ref{tab:cosmo}.
\begin{table}
  \centering
  \begin{tabular}{cccccc} 
  model & $\Omega_{M}$ & $\Omega_V$ & $\Gamma$ &
    $\sigma_{8}$ & h \\ \hline 
  $\Gamma$CDM  & 1   & 0   & 0.25 & 0.64 & 0.5 \\
  OCDM         & 0.3 & 0   & 0.15 & 0.85 & 0.5 \\
  $\Lambda$CDM & 0.3 & 0.7 & 0.15 & 0.85 & 0.5 \\
  \end{tabular}
  \caption{Table showing the parameters of the cosmological models
  adopted in the three N-body simulations.}  \label{tab:cosmo}
\end{table}
Each simulation contained $128^3$ dark matter particles, and particle
positions were output at a large number of epochs (362 for
$\Gamma$CDM, 345 for OCDM \& 499 for $\Lambda$CDM), separated by
approximately equal intervals in time. The box size chosen was
100\,$h^{-1}$Mpc for all three simulations that gave a particle mass
of $2.6\times10^{11}$\msun\ for $\Gamma$CDM and
$7.9\times10^{10}$\msun\ for the other two simulations. Groups of
particles were found using a standard friends-of-friends algorithm
with linking length set to $b=0.2$ times the mean interparticle
separation.

Fig.~\ref{fig:nb_cmf} shows the conditional mass function calculated
for different values of $\Delta\delta_c$. Halos of between 200 and
1500 particles were selected from every output time, corresponding to
masses $1.6\times10^{13}$\msun$<M<3.9\times10^{14}$\msun, and the
particles were traced back to determine the mass of the halo they
resided in at the output time nearest the required
$\Delta\delta_c$. Only halos containing at least 40 particles, and
with over $50\,\%$ of their particles in common with the final halo
were considered to be progenitors. The choice of $50\,\%$ is
reasonably arbitrary, and does not affect the progenitor distribution
significantly. $\sigma_M^2$ was calculated for these halos using the
top-hat filter. We have tried using Gaussian and sharp $k$-space
filters to calculate $\sigma_M^2$ and find that the result is not
significantly affected by this choice.

The values of $\Delta\delta_c$ adopted correspond to $\Delta
z\sim0.01$, $\Delta z=0.1$ and $\Delta z=0.5$ for an Einstein-de
Sitter cosmology. The smallest of these intervals corresponds to
consecutive output from the simulations. 

Comparing Fig.~\ref{fig:psf_cmf} and Fig.~\ref{fig:nb_cmf} provides
weak evidence that the build-up of halos in PS theory is most like
that predicted by the sharp $k$-space filter. This is discussed
further in Section~\ref{sec:conc}.

\section{The cosmological distribution of mergers}  \label{sec:merg}

In order to link halo growth with the observed evolution of the
baryonic component of galaxies, it is important to distinguish between
major mergers, which can disrupt structure creating observable
signatures, and the slow accretion of matter. Where trajectories dip
and rise again (as the filter radius decreases), the mass associated
with the trajectory undergoes a jump. Such an event is shown in
Fig.~\ref{fig:traj} for a trajectory calculated with a sharp $k$-space
filter. These jumps are often associated with mergers, and it is
expected that close congregations of excess overdensity in the linear
density field that initially collapse as single entities and
subsequently merge together, will create such a dip and rise as the
filter increases in size and the number of enclosed clumps
increases. For PS theory with a sharp $k$-space filter, jumps are the
only mechanism for gaining mass as the trajectories are not
differentiable.

\begin{figure}
  \setlength{\epsfxsize}{\columnwidth}
  \centerline{\epsfbox{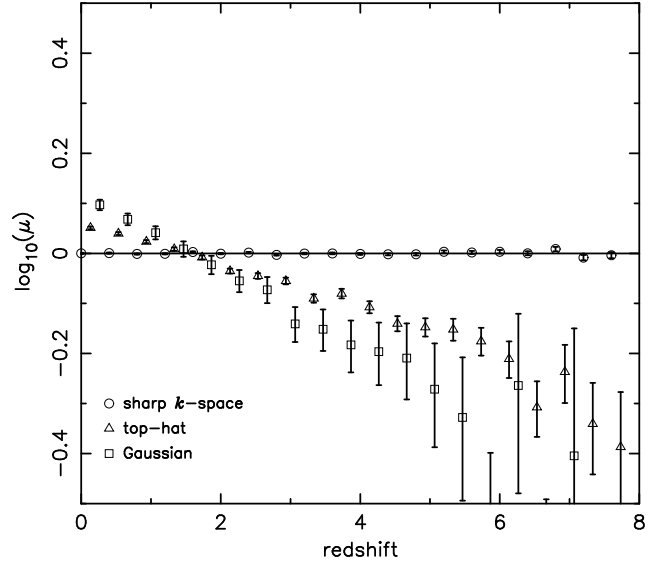}}

  \caption{The normalized proportion of halo creation events that are
  a result of merger, measured from a set of $10^6$ trajectories,
  calculated for each of the three filters described by
  Equations~\ref{eq:fk}~-~\ref{eq:fg}. Here, a merger was defined to
  be a mass jump of at least a quarter of the final mass, with the
  trajectory after the jump also corresponding to a halo with mass
  over a quarter of the final value.}  \label{fig:psf_merg}
\end{figure}

The halo creation events analysed in this Section can be thought of as
being instantaneous. However, numerical limitations are placed on how
well we can follow these events because the trajectories are
discretized in mass and the simulations are discretized in both mass
and time. For the simulations, simultaneous outputs from the
simulations were used to determine whether a halo had just been
created, and whether the creation event was a merger. For the
trajectories, the behaviour of a trajectory before it passed through a
point was analysed to determine whether it was a first upcrossing (a
creation event), and the subsequent behaviour was analysed to
determine the form of creation. Both the trajectories and the
simulations were sufficiently sampled that this numerical problem
should not affect the conclusions drawn from this work. Defining the
merger fraction, $f_{\rm merg}(M_f,z_f)$, by
\begin{equation}
f_{\rm merg}(M_f,z_f)=\left.\frac{\rm No.\ mergers}{\rm No.\ new\ halos}
  \right|_{M_f,z_f},
\end{equation}
where $z_f$ is the final redshift of the halo and $M_f$ is the final
mass of the halo, we can write the normalized proportion of new halos
created by merger as
\begin{equation}
\mu'(M_f,z_f)=\frac{f_{\rm merg}(M_f,z_f)}
  {\bar{f}_{\rm merg}(M_f)}
\end{equation}
where $\bar{f}_{\rm merg}(M_f)$ is the average of $f_{\rm
merg}(M_f,z_f)$ over all redshift. These data were combined over all
final halo masses of interest, weighted by the standard Binomial
variance treating the number of creation events found for each $M_f$
and $z_f$ as the number of `experiments' performed. The resulting
normalized merger fraction, a function of redshift only, is denoted by
$\mu$. This procedure is designed to allow the merger fraction to be a
function of mass and to highlight any cosmological changes in the
overall merger rate.

Fig.~\ref{fig:psf_merg} shows the cosmological variation in $\mu$,
calculated from PS trajectories, for mergers defined such that both
the mass jump and the progenitor correspond to at least a quarter of
the mass of the final halo. Results presented in this Section are not
sensitive to this choice. The numbers of mergers and halo creation
events were determined from a set of $10^6$ trajectories, calculated
as described in Section~\ref{sec:ps} for each of the three filters
given by Equations~\ref{eq:fk}~-~\ref{eq:fg}. The discrete nature of
the trajectories provides the required lower limit of the size of
possible accretion event and stops the number of halo creation events
in any time interval becoming infinite. Data have been combined for
all creation events that result in a halo of mass
$4\times10^{10}\msun<M<10^{13}\msun$, in such a way that the results
should be directly comparable with those calculated from N-body
simulations.

The Markovian nature of the trajectories calculated using a sharp
$k$-space filter means that the probability that a halo was created by
merger is independent of creation time: the normalized merger
probability is flat. However, the trajectories calculated for the
other filters are non-Markovian and the probability of merger is
dependent on cosmological epoch. There is a general trend for these
filters that when a halo is created at early times it's more likely to
have been created via the slow build-up of matter rather than by
merger.

\begin{figure}
  \setlength{\epsfxsize}{\columnwidth} 
  \centerline{\epsfbox{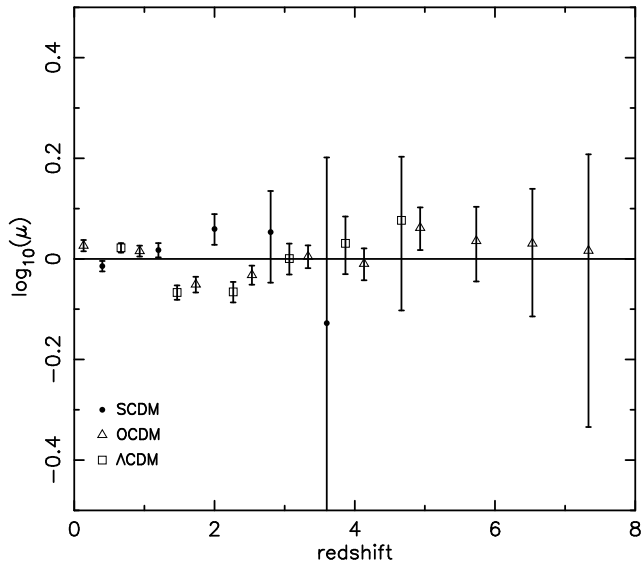}} 

  \caption{The normalized proportion of halo creation events that are
  a result of merger, measured from three N-body simulations from
  which groups of particles were determined using a standard
  friends-of-friends group finder with $b=0.2$. Halo creation was
  defined to have occurred for a given halo at a particular time if at
  least half of the particles had not previously been contained within
  a halo with equal or greater mass. Only halos comprising at least 45
  particles were included in this analysis. A creation event was said
  to have been a merger if there were two possible progenitors at the
  previous output time from the simulation, each with mass greater
  than a quarter of the final halo mass. The data are not strongly
  dependent on this choice, and are consistent with no evolution in
  the proportion of merger events as a means of halo creation.}
  \label{fig:nb_merg}
\end{figure}

It is interesting to compare this behaviour with that observed in
N-body simulations. Here, the discrete nature of the simulations stops
the number of halo creation events within a time interval from tending
to infinity. Halo creation was defined to have occurred for a given
halo at a given time if at least half of the particles had not
previously been contained within a halo with equal or greater
mass. Only halos comprising at least 45 particles were included in
this analysis. A creation event was said to have been a merger if
there were two possible progenitors at the previous output, each with
mass greater than a quarter of the final halo mass. Using the analysis
described above for combining data from halos of different mass, these
data are presented in Fig.~\ref{fig:nb_merg}.

Fig.~\ref{fig:nb_merg} suggests that the distribution of mergers
follows the PS model with a sharp $k$-space filter, with no evidence
for a cosmological trend as predicted by PS theory with either the
top-hat or Gaussian filters. PS theory with the sharp $k$-space filter
would therefore seem to provide the best model for the build-up of
clumps of dark matter determined using the FOF algorithm.

\section{The tie-in with observed galaxy evolution}  \label{sec:obs}

A variety of different techniques have been used to determine the
amount of evolution in the field galaxy merger rate as a function of
redshift: recent studies have used observations of the cosmological
change in either the angular correlation function \cite{neuschaefer},
pair counts \cite{carlberg94,patton} or galaxy morphologies
\cite{lavery}. See Abraham \etal\ \shortcite{abraham98} for a
review. There is now strong evidence that the number of interacting
systems grows rapidly with look-back time. The cosmological change in
the proportion of galaxies involved in mergers is usually fitted by a
power-law of the form $(1+z)^n$, and results are parametrized by the
best fit value of $n$. A recent study using follow-up HST imaging of a
sample of 185 galaxies with measured redshifts from the CFRS and
Autofib-LDSS surveys has provided the most convincing measurement of
$n$ to date \cite{lefevre}. This work extends the pair-count analysis
of Patton \etal\ \shortcite{patton} from $z=0.33$ to $z=0.9$ and
reveals a consistent amount of evolution between such studies and
results from the morphological classification of galaxies. A best fit
value of $n=2.7\pm0.6$ is determined after correction for biases
attributed to selection effects: the combination of ground-based and
HST observations means that the sample is biased towards finding close
pairs that are thought to be single galaxies in the magnitude limited
ground-based sample, but which can be resolved into separate galaxies
by HST.

These observational studies select galaxies based on properties of the
baryonic galaxy component. In order to model such observations, the
predicted evolution in the halo merger rate has to be combined with
the time-scale over which the tracers of halo merger are visible,
including any time difference between the merger of the baryonic
components and the halos caused by the dynamical friction
time-scale. The dynamical friction time-scale describes how long it
takes for a galaxy (or progenitor) to spiral to the centre of a
halo. If this takes a long time compared to the cosmological
time-scale of interest, it would affect the relative distribution of
baryonic and halo mergers. However, by concentrating on galaxies in
small groups or field galaxies at relatively low redshifts, the
dynamical friction timescale is expected to be small
\cite{barnes92}. For such studies, or by using galaxy counts rather
than morphologies, the observed evolution is expected to be driven by
the proportion of halos that have undergone recent merger.

Percival \etal\ \shortcite{ev2} showed that multiplying the mass
function by $d\delta_c/dt$ leads to the joint distribution of halo
creation events in mass and time. In Section~\ref{sec:merg} the
distribution of mergers in numerical simulations was shown to follow
that predicted by PS theory with a sharp $k$-space filter, and to have
the same form as the distribution of creation events. The proportion
of halos that have just undergone merger is given by the creation rate
divided by the number of halos, given by the standard PS mass function
(Equation~\ref{eq:psmf}). The resulting cosmological change is
proportional to $d\delta_c/dt$, and only depends on the Friedmann
equation. A method for calculating $d\delta_c/dt$ was given by
Percival \etal\ \shortcite{ev2}. Because PS theory does not include
sub-halos that form part of a larger collapsed system, this is only
valid for field galaxies. This will also be the evolution of any
combination of halos of different mass, provided the combination is
independent of redshift. For the three cosmologies parametrized in
Table~\ref{tab:cosmo}, approximating $d\delta_c/dt$ by $(1+z)^n$ gives
$n=2.5$ for SCDM, $n=2.1$ for OCDM and $n=2.3$ for $\Lambda$CDM. These
values are in agreement with the best fit value of Le~F\`{e}vre \etal\
\shortcite{lefevre}, $n=2.7\pm0.6$ within the errors and
uncertainties. Further observations to provide more accurate
observational data could, in principle, provide an interesting
mechanism for constraining the best-fit cosmology.

\section{Conclusions}  \label{sec:conc}

The build-up of halos predicted by PS theory has been shown to be
strongly dependent on the filter used to smooth the density
field. None of the filters provides a perfect match for the build-up
of structure in the numerical simulations presented in this paper. The
simulation mass function \cite{sheth,ev2} is most similar with PS
theory with a top-hat filter, as is the halo creation rate.

The top-hat and Gaussian filters predict mass growth due to continuous
accretion with the occasional merger, while the sharp $k$ space filter
predicts mass growth completely due to mergers. Because of this, the
predicted backwards conditional mass function is strongly dependent on
the choice of filter. Analysing numerical simulations suggests a
backwards conditional mass function most like that of PS theory with
the sharp $k$-space filter, although again, the match is not
perfect. The resolution of the simulations means that it is difficult
to resolve mergers with small enough halos. However, there is weak
evidence that the simulations contain more mergers with small mass
halos than are predicted by the smoother trajectories associated with
the top-hat and Gaussian filters.

For the sharp $k$-space filter, the cosmological distribution of halo
mergers is independent of merger parameters. This result follows from
the Markovian nature of the trajectories produced. For the other
filters, which predict non-Markovian trajectories, the probability of
merger is shown to be dependent on the halo creation epoch. No
evolution is seen in the probability that a given halo was created by
merger measured from N-body simulations, and PS theory with a sharp
$k$-space therefore provides the best fit to these data. This provides
the most convincing evidence that PS theory with a sharp $k$-space
filter provides the best model of halo growth. In this work, a
standard FOF group finder was used. An interesting test would be to
analyse the observed build-up of halos using a number of different
algorithms to find halos. However, this is not expected to
significantly alter the conclusions because the build-up of mass in
the simulations is independent of the group finder.

When combined with the link between the mass function and halo
creation rate described in Percival \etal\ \shortcite{ev2}, this leads
to the interesting result that the cosmological change in the
proportion of halos that have undergone recent merger is predicted to
be independent of power spectrum, halo mass, mass function, and to
only depend on the Friedmann cosmology through $d\delta_c/dt$. In
Section~\ref{sec:obs} it was argued that the field galaxy merger rate
should be driven by the halo merger rate, and that observations of
field galaxies estimate the cosmological evolution in the halo merger
rate. In Section~\ref{sec:obs}, recent observational data were shown
to be consistent with the presently favoured $\Lambda$CDM cosmology,
although they are insufficiently accurate to usefully distinguish
between cosmological parameters. It is hoped that future studies of
this evolution will provide sufficiently accurate results that this
picture will change.

\section{Acknowledgements}
I would like to thank John Peacock and the anonymous referee for
useful comments that helped to improve this paper. The Hydra N-body
code \cite{couchman} was kindly made available by the Hydra
consortium.

\end{document}